# Field-induced length changes in the spin-liquid candidate κ-(BEDT-TTF)$_2$Cu$_2$(CN)$_3$

Rudra Sekhar Manna[1], Mariano de Souza[1,2], John A. Schlueter[3], and Michael Lang[*,1]

[1] Physics Institute, Goethe-University Frankfurt(M), Max-von-Laue Str. 1, D-60438 Frankfurt(M), Germany
[2] Departamento de Física, IGCE, Unesp - Universidade Estadual Paulista, Caixa Postal 178, CEP 13500-970, Rio Claro (SP), Brazil
[3] Materials Science Division, Argonne National Laboratory, Argonne, IL 60439, USA

* Corresponding author: e-mail milang@physik.uni-frankfurt.de, Phone: + 49 69798 47241, Fax: + 49 69798 47250

Measurements of the coefficient of thermal expansion on the spin-liquid candidate κ-(BEDT-TTF)$_2$Cu$_2$(CN)$_3$ have revealed distinct and strongly anisotropic lattice effects around 6 K – a possible spin liquid instability. In order to study the effects of a magnetic field on the low-temperature spin-liquid state, dilatometric measurements have been conducted both as a function of temperature at $B = const.$ and as a function of field at $T = const.$ While the 6 K anomaly is found to be insensitive to magnetic fields $B \leq 10$ T, the maximum field applied, surprisingly strong $B$-induced effects are observed for magnetic fields applied along the in-plane $b$-axis. Above a threshold field of $0.5$ T $< B_c \leq 1$ T, a jump-like anomaly is observed in the $b$-axis lattice parameter. This anomaly, which is located at 8.7 K at $B = 1$ T, grows in size and shifts to lower temperatures with increasing the magnetic field. Although the anomaly bears resemblance to a first-order phase transition, the lack of hysteresis suggests otherwise.

**1 Introduction** The title organic charge-transfer salt is a layered Mott insulator where spin-1/2 dimers reside on a weakly anisotropic triangular lattice. The material is considered as a prime candidate for the realization of a quantum spin liquid (QSL) state due to the lack of long-range magnetic order down to 20 mK [1,2] despite the sizable magnetic exchange coupling constant of $J/k_B \sim 250$ K. More recent investigations, aiming at identifying the nature of the QSL and its potential instabilities, have raised a number of important questions. Points at issue are whether or not the QSL state has a gap [3,4] and the nature of the mysterious 6 K anomaly. The latter feature manifests itself in distinct anomalies in various quantities such as $^{13}$C NMR [5], specific heat [3], thermal conductivity [4], magnetic susceptibility [6] and thermal expansion [6]. The pronounced lattice effects at 6 K prove the thermodynamic nature of the anomaly and provide evidence for a second-order phase transition [6]. Furthermore, the thermodynamic analysis in [6] showed that spin degrees of freedom alone cannot account for the observed entropy release, indicating that charge degrees of freedom are involved in the transition. According to recent muon spin rotation experiments [7], the QSL state in the present material is unstable in a magnetic field. It has been found that the application of a tiny field as small as a few mT suffices to induce a quantum phase transition from the QSL state into an antiferromagnetic phase with strongly suppressed moment. At fields of around 4 T, a second quantum phase transition was found, suggesting a deconfinement of the spin excitations. These observations call for further experimental and theoretical studies, in particular on the field-dependent effects, to unravel the nature and the exotic properties of the QSL state realized in this molecular material.

Here we report thermal expansion measurements performed in varying magnetic fields. For fields aligned along the in-plane *b*-axis, pronounced step-like anomalies were observed in the *b*-axis lattice parameter.

**2 Experimental** The thermal expansion measurements were carried out by using an ultrahigh-resolution capacitive dilatometer, built after [8], enabling the detection of length changes $\Delta l \geq 10^{-2}$ Å. The single crystals were prepared by following the standard procedure [9]. For the present study, the measurements were performed on the same single crystal as used in Ref. [6].

**3 Results and Discussion** Figure 1 shows relative length changes $\Delta l_b(T)/l_b = [l(T) - l(T_0)]/l(T_0)$, with $T_0 = 300$ K, measured along the in-plane *b*-axis as a function of temperature in varying magnetic fields $B \parallel b$. The data have been taken upon cooling at a rate of -1.5 K/hour. The results at zero field show a broad minimum centred around 8 K, corresponding to a change of sign in the *b*-axis thermal expansion coefficient $\alpha_b = l_b^{-1} \partial l_b / \partial T$, cf. Refs. [6,12]. On the scale used in figure 1, the abrupt break in the slope in $\Delta l_b/l_b$ near 6 K (arrow), giving rise to a sharp minimum in $\alpha_b$ (cf. Fig. 2 in [6]), cannot be discerned. By the application of a magnetic field of 1 T, a discontinuity in $\Delta l_b/l_b$ at 8.7 K becomes visible. This discontinuity grows in size and progressively shifts to lower temperatures with increasing the field. A peculiarity of this *B*-induced anomaly is the undershoot (overshoot) behaviour on the high (low)-temperature side of the discontinuity.

Abrupt length changes are generally considered to be a signature of a first-order phase transition. Other characteristics of a first-order transition include (i) hysteretic behaviour upon cooling and warming and (ii) different slopes in $\Delta l_b/l_b$, i.e. different thermal expansion coefficients $\alpha_b$, for temperatures above and below the transition, see, e.g. Ref. [10] for thermal expansion data across the first-order Mott transition. For the present case, measurements upon cooling and warming at a rate of ±1.5 K/h failed to resolve any significant hysteresis. As an example, Fig. 1 shows data for a cooling and warming run at $B = 8$ T. Moreover, $\alpha_b$ remains virtually unchanged upon crossing the anomaly, except in the immediate vicinity of the discontinuity (not shown, [11]). Both observations point against an interpretation in terms of a first-order transition.

A further observation, highlighted in figure 2, may also suggest an interpretation other than a first-order transition. We show, for example, the 6 T data together with the zero-field curve without employing a vertical shift. Interestingly, the data collapse both at the high and low-*T* end, but deviate from each other at intermediate temperatures. This demonstrates that the total change of the *b*-axis lattice parameter on cooling from 12 K down to 4.5 K is virtually unaffected by the presence of a magnetic field of 6 T. The same observation holds true, within the uncertainty of the experiment, for all other $B = const.$ curves shown in figure 1. The gradually growing departure of the $B = const.$ data from the zero-field curve upon cooling, i.e. an additional *B*-induced lattice strain, is released by the jump-like anomaly.

The absence of any observable discontinuity in the low-temperature $\Delta l_b/l_b$ data at zero field and 0.5 T suggests that a finite field is required to induce the anomaly. From the present data, a threshold field $B_c$ can be estimated to $0.5\,\text{T} < B_c \leq 1\,\text{T}$.

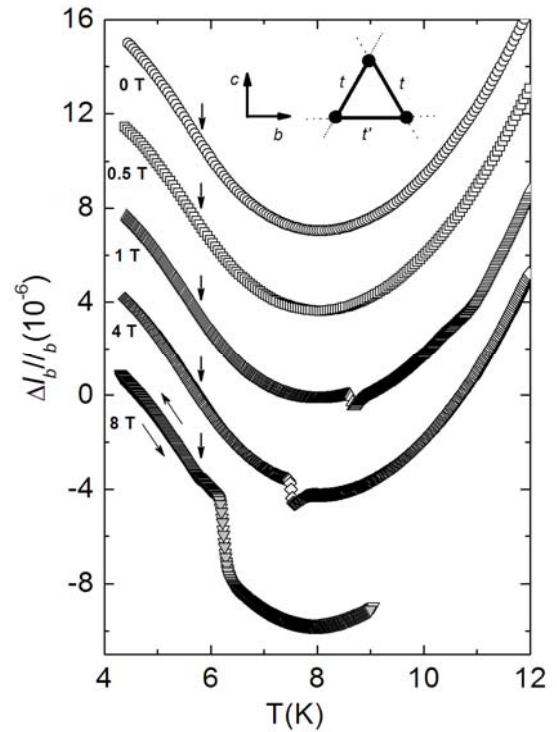

**Figure 1** Relative length changes of single crystalline κ-(BEDT-TTF)$_2$Cu$_2$(CN)$_3$ measured along the in-plane *b*-axis as a function of temperature for varying magnetic fields $B \parallel b$. The data have been shifted vertically for clarity. The arrows indicate the position of a break in the slope, corresponding to a second-order phase transition anomaly in the coefficient of thermal expansion, cf. Refs. [6,12]. For $B = 8$ T, data taken upon cooling and warming are shown.

In order to map out the position of the *B*-induced anomaly in the temperature-field plane, magnetostriction measurements at varying constant temperatures have been carried out. These measurements, which will be the subject of an independent publication [11], reveal consistent results to those obtained in the temperature-dependent runs.

In figure 3, we compile the positions of both field-induced anomalies as derived from measurements at $B = const.$ (open circles) and $T = const.$ (closed triangles). The location of the anomaly is not too far from the phase transition line between a quantum critical region and the weak-moment antiferromagnetic high-temperature phase (WAF$_H$) drawn in Ref. [7] on the basis of their muon spin rotation experiments performed on polycrystalline material. In order to make contact to those observations, however, it is important to note that the $B$-induced anomaly, reported in the present work, is observed only for fields aligned along the in-plane $b$-axis. No such field-induced anomalies were found for fields aligned along the $a$- and $c$-axis [12]. In addition, we stress that on the basis of the present data, the question of whether or not the $B$-induced anomaly manifests a phase transition cannot be definitely answered.

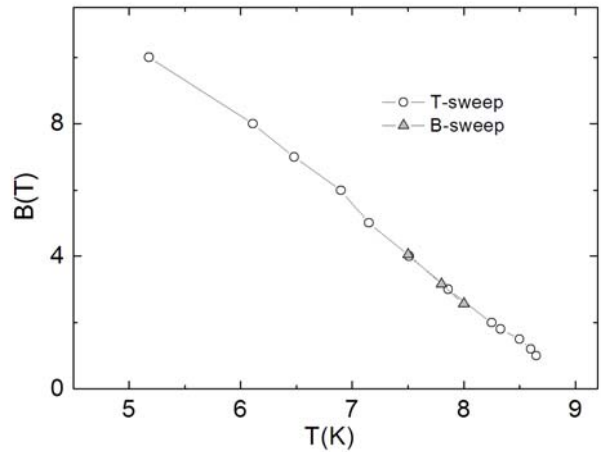

**Figure 3** Position of the $B$-induced anomalies in $\Delta l_b/l_b$ for $B$ applied parallel to the $b$-axis as determined from measurements as a function of $T$ at $B = const.$ (open circles) and as a function of $B$ at $T = const.$ (closed triangles).

**Acknowledgements** Work at the Goethe-University Frankfurt(M) has been supported by the German Science Foundation via the SFB/TR49. Work supported by Argonne, a U.S. Department of Energy Office of Science laboratory, operated under Contract No. DE-AC02-06CH11357. We thank T. Sasaki for fruitful discussions.

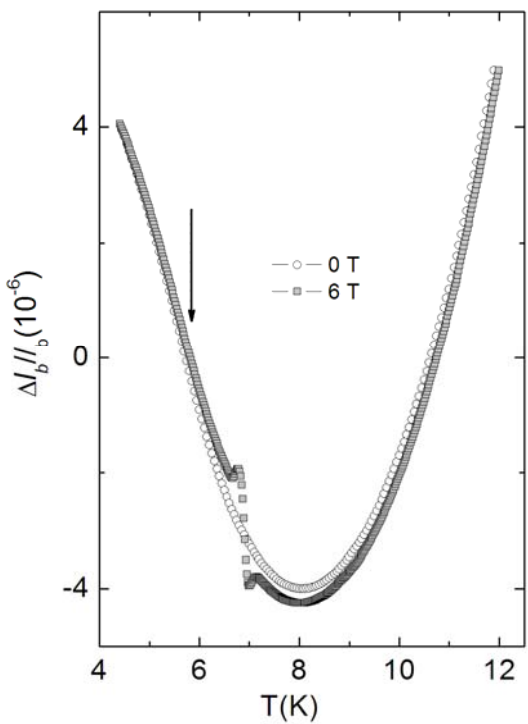

**Figure 2** Plot of the $\Delta l_b/l_b$ data at $B = 0$ and 6 T from figure 1 without employing a vertical shift. The arrow indicates the position of a break in the slope of the zero-field data, corresponding to a second-order phase transition anomaly in the coefficient of thermal expansion, at around 6 K. The 6 K transition remains unchanged in the data taken at 6 T.